\documentclass[11pt]{article}
\usepackage{moriond,epsfig}

\def\Journal#1#2#3#4{{#1} {\bf #2}, #3 (#4)}

\def\NIMA{{\em Nucl. Instrum. Methods} A}

\def\PLB{{\em Phys. Lett.}  B}
\def\PRL{\em Phys. Rev. Lett.}
\def\PRD{{\em Phys. Rev.} D}

\def\EPJ{{\em Eur. Phys. J.} C}

\def\be{\begin{equation}}
\def\ee{\end{equation}}
\def\bea{\begin{eqnarray}}
\def\eea{\end{eqnarray}}

\def\kpppc{K^{\pm}\rightarrow\pi^{\pm}\pi^+\pi^-}
\def\kpppp{K^{+}\rightarrow\pi^{+}\pi^+\pi^-}
\def\kpppm{K^{-}\rightarrow\pi^{-}\pi^-\pi^+}
\def\kpppn{K^{\pm}\rightarrow\pi^{\pm}\pi^0\pi^0}
\def\pppc{\pi^{\pm}\pi^+\pi^-}
\def\pppn{\pi^{\pm}\pi^0\pi^0}
\def\ag{A_g}
\begin{document}
\vspace*{4cm}
\title{SEARCH FOR DIRECT CP VIOLATION IN $\kpppc$ DECAYS BY 
NA48/2~\footnote{Talk given at XXXXth Rencontres de Moriond, 
ElectroWeak Interactions and Unified Theories, La Thuile, March 2005.}
}

\author{ I. Mikulec~\footnote{
On behalf of the NA48/2 collaboration.} }

\address{\"Osterreichische Akademie der Wissenschaften, Institut f\"ur
  Hochenergiephysik, A-1050 Wien, Austria}

\maketitle\abstracts{
First preliminary measurement of the direct CP-violating parameter $\ag$ by the
NA48/2 experiment at CERN SPS is presented. Using more than 1.6
billions of charged kaon decays into three charged pions, the charge
asymmetry in the $\kpppc$ Dalitz plot slope, $\ag$, has been measured to
$\ag = (0.5 \pm 3.8) \times 10^{-4}$. This result is more than an order
of magnitude more precise than results of previous experiments.
}

\section{Introduction}

Violation of the CP symmetry, and especially direct CP violation
in the decay amplitude, due to its subtle nature, is an important
window into physics beyond Standard Model (SM). 

It took more than three decades since the discovery of CP violation in
the neutral kaons by Christenson, Cronin, 
Fitch and Turlay~\cite{cro}, until direct CP violation was definitively
established. After an unconfirmed indication by NA31~\cite{na31}, 
KTeV and NA48 have
demonstrated in the late 90's with high significance that direct CP
violation exists in the decays of neutral kaons into two
pions~\cite{epc,epk}. In the year 2001, B-factory
experiments Babar and Belle have found CP violation in the system of
neutral B mesons~\cite{bb1} and last year also the direct CP violation in
B-decays has been demonstrated~\cite{bb2}.

In order to explore possible non-SM enhancements to heavy-quark loops
which are at the core of direct CP-violating processes, all
manifestations of direct CP violation need to be experimentally
studied and measured. In kaons, besides the already measured parameter
$\varepsilon'$ in $K_L \rightarrow \pi\pi$ decays, the most promising
complementary observables are decay rates of GIM suppressed 
rare kaon decays~\cite{aug} which proceed through flavor-changing neutral
currents and the asymmetry between $K^+$ and $K^-$ decays into three
pions.

%\newpage

The $K^{\pm}\rightarrow 3\pi$ matrix element is usually described in
a polynomial expansion of two Dalitz variables $u$ and $v$:
\be
|M(u,v)|^2 \propto 1 + gu + hu^2 + kv^2 + O(u^3,v^3)
\ee
where $|h|,|k| \ll |g|$ and
\be
u=\frac{s_3-s_0}{m_{\pi}^2} \hspace{1cm} \mathrm{and} 
\hspace{1cm} v=\frac{s_2-s_1}{m_{\pi}^2} 
\ee
with $s_i=(p_K-p_{\pi i})^2$ and $s_0=\sum{s_i}/3 \
(i=1,2,3)$. Index $i=3$ stands here for the odd pion~\footnote{
Other two, even, pions have equal charges.}.
Slope parameters $g$ can differ 
between the $K^+$ and $K^-$
only due to direct CP violation~\footnote{Due to absence of
  mixing only direct CP violation is possible in decays of charged
  kaons.}. SM predictions for the corresponding asymmetry
\be
\ag = \frac{g^+ - g^-}{g^+ + g^-}
\ee
vary between few~$10^{-6}$ to few~$10^{-5}$~\cite{tsm}.
An analogous asymmetry of integrated decay rates, $A_{\Gamma}$, is
expected to be more than an order of magnitude smaller.
Several experiments~\cite{exp} have searched for the asymmetry $\ag$. The
precision reached in both $\pppc$ and $\pppn$ decay modes so far
is at the level of few~$10^{-3}$.
Existing theoretical calculations
involving processes beyond SM~\cite{tbs} predict
substantial enhancements of the asymmetry 
$\ag$ partially within reach
of the NA48/2 experiment.

NA48/2 is an extension of the original
experimental program of the experiment NA48 at the 
CERN SPS which has successfully
accomplished the main goal to establish and measure the direct CP
violation in the decays of neutral kaons into two
pions~\cite{epc}. The primary aim of the NA48/2 extension~\cite{pro} 
is to measure the
parameter $\ag$ in $\pppc$ and
$\pppn$ modes to $\sim 2-4\times10^{-4}$.

NA48/2 took data in two periods. In total, during the 50 day run in 
the year 2003 and
the 60 day run in the year 2004 about 4 billions of $\kpppc$ and 
about 200 millions of
$\kpppn$ decays have been collected and taped. The total recorded data
volume amounts to about 200 TB. This paper describes the analysis and the
preliminary result based on about 1.6 billions of $\kpppc$ decays 
taken during the first 2003 run.

\section{Description of the Experiment}

NA48/2 deploys a novel system of two
simultaneous charged beams with opposite charges. This allows to record
decays of $K^+$ and $K^-$ at the same time achieving a significant
cancellation of systematic effects in the charge asymmetry measurement.

\begin{figure}
\psfig{figure=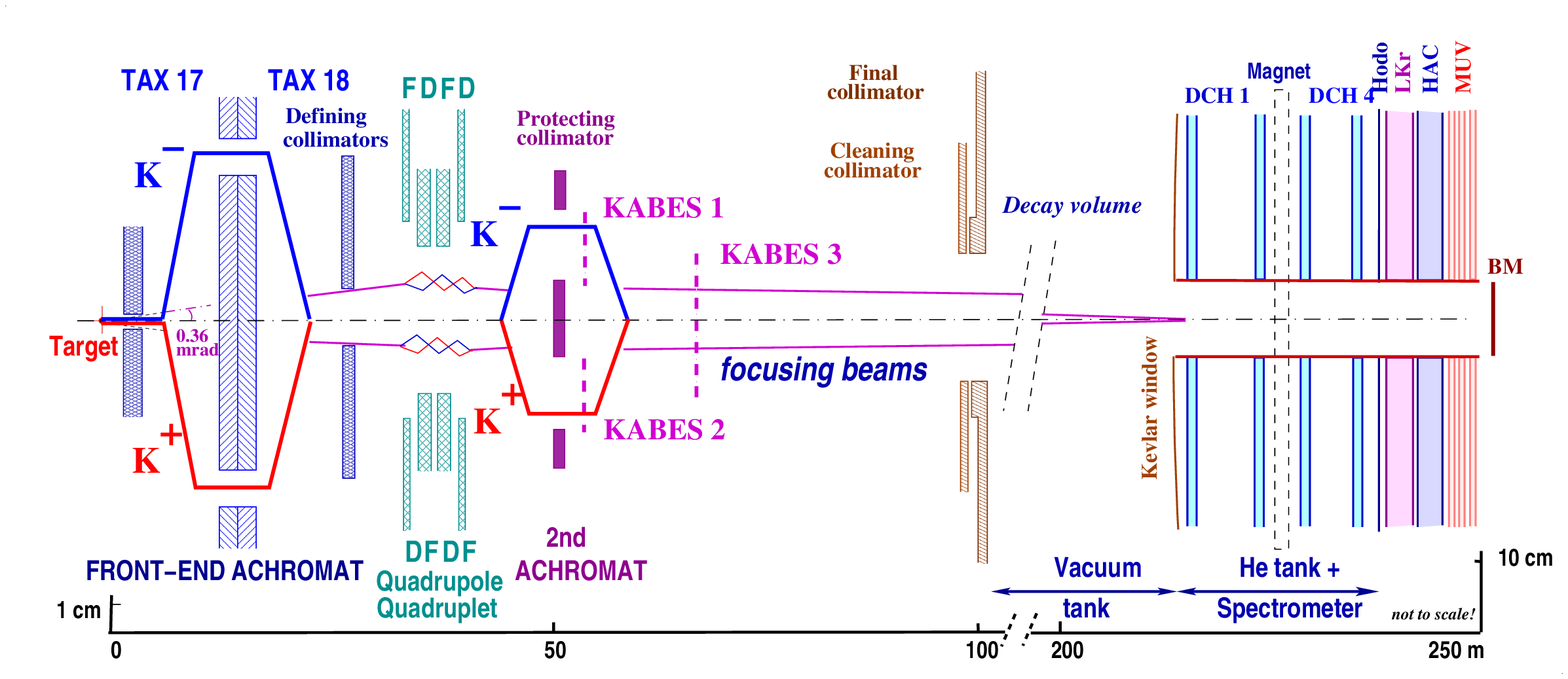,width=\textwidth}
\caption{Lateral view of the NA48/2 beam and detector. The vertical
  scale is strongly enhanced.
\label{fig:beams}}
\end{figure}

The beams (Fig.~\ref{fig:beams})
of charged particles are derived from protons
from the SPS which impinge with an intensity of about
$7\times 10^{11}$ protons per 5 s pulse on a beryllium target of 2
mm diameter and 40 cm length at a zero angle of incidence. The central
momentum of 60 GeV and the momentum bite of $\pm$3 GeV are then selected
symmetrically for positively and negatively charged particles in the
first achromat unit which splits the two beams in the vertical plane
and recombines them on the same axis. 
Then both beams pass through a series of focusing
quadrupoles and are again split and cleaned in the second
achromat. The second achromat together with the KABES
detector~\cite{kab} serves also as a beam spectrometer. The $K^+/K^-$ 
production ratio is about 1.8 (the analysis is independent of this ratio).

After the second achromat both beams follow the same path. After
passing the cleaning and the final collimators they
traverse the entire $\sim$114 m long decay volume superimposed with a
precision of about a millimeter. This superposition symmetrises the
acceptances and contributes to the reduction of systematic biases. 
The decay region is comprised in a
vacuum tank. The whole detector apparatus downstream the decay volume is
almost identical to the NA48 experimental setup.

The detector used for the reconstruction of $\kpppc$ decays is the
magnetic spectrometer. The spectrometer is housed in a tank filled
with helium gas at atmospheric pressure separated from the vacuum tank
by a KEVLAR window. A thin evacuated beam tube, traversing the centre
of the detector,
allows the undecayed beam  particles to continue in vacuum. Two drift
chambers are located before and two after the central dipole magnet
which induces a horizontal transverse momentum kick of about 120 MeV/c
to all charged particles. 
The drift chambers have an octagonal shape with an area of about 4.5
m$^2$. Each is made of four sets of two staggered
planes of sense wires oriented along four 45$^\circ$ directions. The
momentum resolution of the magnetic spectrometer 
is $\sigma(p)/p = 1.0\% \oplus 0.044\% p$ with $p$ in
GeV/c units.

The $\kpppc$ decays are triggered with a two-level trigger system. At
the first level, the rate is reduced to few hundreds of kHz by
requiring at least two hits in a scintillator hodoscope placed behind
the magnetic spectrometer. The second level trigger, which consists of
hardware coordinate builders and a farm of asynchronous
microprocessors, reconstructs tracks using data from the drift
chambers. At least two tracks are required to converge within 5 cm in
the decay volume. In
the majority of the analysed data, level one triggers rejected by this
condition are examined further and
accepted even if only one track is reconstructed by
the second level trigger. In this case it is required that this track is
kinematically incompatible
with a $\pi^{\pm}\pi^0$ decay, assuming the decaying kaon had momentum of 60
GeV/c and was moving along the beam axis. The resulting trigger rate
is about 10 kHz.

The description of other components of the NA48 detector apparatus can
be found elsewhere~\cite{epc}.

\section{Measurement Method}

The measurement is based on comparing the $u$-distributions of $K^+$ and
$K^-$ decays. In case of the $\pppc$ final state, given the actual 
value~\cite{pdg} of
$g=(-0.2154 \pm 0.0035)$, the ratio $N_{K+}(u)/N_{K-}(u)$ is
proportional with sufficient precision to ($1+\Delta g u$). $\ag =
\Delta g/2g$ is extracted from a linear fit to the ratio $N_{K+}(u)/N_{K-}(u)$.

The presence of magnets both in the beam sector (achromats, focusing
quadrupoles, etc.) and in magnetic spectrometer introduces an
unavoidable charge asymmetry of the apparatus. In order to equalise
local effects on $K^+$ and $K^-$ beams the achromat and quadrupole
polarities were reversed during data taking on an approximately 
weekly basis. The polarity of the spectrometer magnet has been
reversed every day~\footnote{During the data taking in the
  year 2004 the period of the spectrometer magnet reversal has been
decreased to few hours.}. The whole approximately two-week cycle
represents a super-sample which is treated in the analysis as an
independent data unit. In the period of 2003, four super-samples 
have been collected.

Each super-sample contains four $\kpppp$ and four $\kpppm$
samples with different combination of 
achromat and spectrometer magnet polarities. The ratio
$R(u)$ is obtained as a product of four $N_{K+}(u)/N_{K-}(u)$ ratios:
\be
\label{eq:quadr}
R(u) = R_{US} R_{UJ} R_{DS} R_{DJ} \approx \overline{R} (1+4\Delta g u)
\ee
where $U$ represents a configuration in which $K^+$ beam runs through
the upper beam path in the achromats and $D$
the lower. The index $S$ represents spectrometer magnet polarity in
which decay products having the same charge as the corresponding beam are
deflected to the right with respect to the direction of the
beam (towards the Saleve mountain) and $J$ to the left (towards the
Jura mountain). A linear fit to Eq.~\ref{eq:quadr} results in two parameters,
normalisation $\overline{R}$ and $\Delta g$ from which
$\ag$ is extracted.

The quadruple ratio technique in Eq.~\ref{eq:quadr} completes the
procedure of magnet polarity reversal. It allows a three-fold cancellation of
systematic biases:
\begin{itemize}
\item beam line local biases cancel between $K^+$ and $K^-$ samples in
  which the beam follows the same path;
\item local detector biases cancel between $K^+$ and $K^-$ samples
  deflected toward the same parts of the detector;
\item as a consequence of simultaneous beams, global time-variable
  biases cancel between $K^+$ and $K^-$ samples.
\end{itemize}
This method is independent on the relative size of the samples with
different magnet configurations. On the other hand, the statistical
uncertainty depends on the statistically weakest of the eight samples
involved in Eq.~\ref{eq:quadr}. Further reduction of systematic biases
especially due to presence of stray permanent magnetic 
fields (earth's field, vacuum tank
magnetisation) is obtained by maintaining azimuthal symmetry in the
acceptance. 

Using the method described in this section, the result remains
sensitive only to time variation of asymmetries in experimental
conditions which have a characteristic time smaller than corresponding
field-alternation period. 

Due to superposition of the two beams the
measurement does not need a Monte Carlo calculation of the
acceptance. Nevertheless, detailed GEANT-based~\cite{gea} 
Monte Carlo simulation
has been developed as a tool for systematic studies. The Monte Carlo 
simulation includes full detector geometry and material description,
simulation of time-variable local drift chamber inefficiencies and
time-variations of the beam geometry and drift chamber alignment.

\section{Data Analysis}

Several stages of compaction and filtering were necessary in order to
reduce the data to a size suitable for the analysis. 
At least three reconstructed tracks in
magnetic spectrometer, loose acceptance and quality cuts as well as at
least one
good reconstructed three-track vertex are required in the
pre-selection phase.

Tracks are reconstructed from hits in drift chambers using
the measured magnetic field map rescaled according to the recorded current
in the analysing magnet of the spectrometer. Chambers and individual
wires are aligned
using data collected in special runs in which muons were recorded 
with spectrometer magnet off.

The three-track vertex is reconstructed using track segments from
the first half of the spectrometer correcting the extrapolations for the small 
magnetic fields due to magnetisation of the vacuum tank and from the
earth's field. Using measured momenta and track directions obtained at
the vertex the invariant mass of three pions is calculated.
The stray-field correction, calculated based on a
three-dimensional field map measured in the entire vacuum tank,
reduces the azimuthal variation of the
reconstructed invariant mass of $\sim$1 MeV/c$^2$ by an order of magnitude. 

The invariant mass resolution is about 1.7 MeV/c$^2$. Since the decay $\kpppc$
is the dominant three-track decay, the sample is background free. The
tails of the invariant mass distribution are dominated by
events in which one of the three pions decayed and the spectrometer
reconstructs the track of the resulting muon~\footnote{In order to avoid
charge-dependent biases, the muon tracks are not vetoed by the NA48 muon-veto
system.}. Using Monte Carlo simulations, the tails due to pion decays
have been shown to be highly symmetric between $K^+$ and $K^-$
samples. Only far tails are rejected by the 
cut $|m_{\pi\pi\pi}-m_{K}| < 9$~MeV/c$^2$, where $m_K$ is the
PDG value of the charged kaon mass~\cite{pdg}. The systematic
uncertainty due to pion decays, 
limited by the statistical precision of the generated Monte Carlo
samples, is $\delta(\Delta g)=0.4\times 10^{-4}$.

The most important feature determining the acceptance is the beam
tube traversing the centre of all drift chambers. In order to securely
exclude the central insensitive areas, all three tracks are required to
cross the first drift chamber at least 11.5 cm from the beam centre
and the last drift chamber at least 13.5 cm. The latter cut takes into
account the additional beam deflection of about 2 cm with respect to the centre
of the last drift chamber due to the
spectrometer magnet. This important acceptance cut is related to the
beam centre rather than to the centre of the detector. The reason is,
that the beam optics can control the mean beam position only to about
$\pm$1 mm. The actual beam position is continuously monitored to a
much better precision by 
calculating the momentum-weighted centre of gravity 
of three pions at first and
at the last drift chamber planes, independently for $K^+$ and $K^-$. In addition to
the time variation of the beam position, also the dependence of the
beam position on the kaon momentum ($\sim\pm$1 mm in
horizontal and $\sim\pm$1 cm in vertical direction) is taken into
account. In this way the $K^+$ and $K^-$ acceptances 
cancel entirely and no Monte Carlo calculation is needed to correct for their
difference. 
A conservative limit on residual systematic uncertainty,
$\delta(\Delta g)=0.5\times 10^{-4}$, was determined by studying the
sensitivity to various acceptance definitions.

The measurement of the pion momenta is based on the knowledge of the
magnetic field in the spectrometer magnet and on the tracking
information from the drift chambers. The relative variations of the 
current in the magnet can be controlled down to about
$5\times10^{-4}$. Smaller variations are continuously 
corrected by forcing the mean
reconstructed $\pi\pi\pi$ mass to the PDG kaon mass~\cite{pdg} 
with relative precision of
about $10^{-5}$. This is done 
by scaling the measured track momenta symmetrically for positively and
negatively charged tracks. As this effect is charge symmetric, 
by collecting $K^+$ and $K^-$ simultaneously it cancels in the ratio 
 $R(u)$.

A difference between the reconstructed $\pi\pi\pi$ invariant masses
of $K^+$ and $K^-$ is an unambiguous measure of a residual 
horizontal drift chamber misalignment. An uncorrected 
horizontal shift of the chambers can lead to a charge-antisymmetric
mis-measurement of the momenta.
Observing this invariant-mass difference as
a function of time, revealed significant, up to about 0.2 mm,
movements of the drift chambers between individual alignment runs. 
Correcting the measured momenta by $p'=p(1+q\beta p)$, where $q$ is
the sign of the track charge and $\beta$ is proportional to the
difference of the invariant mass between $K^+$ and
$K^-$~\footnote{A shift of last drift chamber by 1 $\mu$m 
corresponds to about 1.5 keV change in
the invariant mass.}, 
reduces the effect to a practically
negligible level~\footnote{In the 2004 data-taking the alignment runs
  were performed on a more frequent, weekly, basis.}. 

A potential source of systematic bias is the trigger. Inefficiencies
of different trigger components are studied and 
measured using control samples from
low bias triggers collected along with the main triggers. 
The
rate-dependent parts of inefficiencies are assumed to be charge symmetric
(simultaneous beams). 
The
inefficiency of the first level was measured to be small, about
$7\times10^{-4}$ and stable in time. No correction is applied and an
uncertainty of $\delta(\Delta g)=0.4\times 10^{-4}$, limited by the
statistics of the control sample, has been attributed to this
part. The rate independent 
inefficiency of second level trigger varies with time between 0.2 and
1.8\%. The main source of this inefficiency are local drift chamber
inefficiencies which are more important in the trigger than in the
reconstruction due to reduced redundancy. All samples 
are corrected by the measured 
second level trigger efficiencies as a function of $u$. The 
uncertainty is fully dominated by the statistics of the control
samples. 

Further sources of systematic effects were studied and evaluated. Bias
due to resolution and the calculation of $u$
was studied by changing the way $u$-variable is calculated from
measured track momenta. Using odd or even pion tracks changes the
resolution as a function of the position in the Dalitz plot ($u$,
$v$). Additional studies were performed excluding various parts of 
the Dalitz plot. The preliminary result quoted in this paper is
obtained from the fit restricted in the $u$-interval between $-1$ and
$+1$. Effects due to pile-up of signals from different kaons due to
high instantaneous intensity were studied by comparing samples with
different amounts of recorded extra tracks close in time to the event.
Charge-asymmetric material effects have been found negligible by
studying amount and composition of the material in front of and in the
chambers and taking into account pion spectra. Track charge
misidentification was evaluated from events with three tracks of equal
charge. Table~\ref{tab:syst} summarises all systematic uncertainties
attributed to the result.

\begin{table}[t]
\caption{Preliminary limits 
on systematic and trigger uncertainties on $\Delta g$ in units of
  $10^{-4}$. \label{tab:syst}}
\vspace{0.4cm}
\begin{center}
\begin{tabular}{|l|r|}
\hline
Acceptance and beam geometry    & 0.5 \\
Spectrometer alignment          & 0.1 \\
Analysing magnet                & 0.1 \\
Pion decay                      & 0.4 \\
Calculation of $u$ and fitting  & 0.5 \\
Pile-up                         & 0.3 \\ \hline
Total systematic uncertainty    & 0.9 \\ \hline \hline
Trigger efficiency - level 1    & 0.4 \\
Trigger efficiency - level 2    & 0.8 \\ \hline
Total trigger uncertainty       & 0.9 \\
\hline
\end{tabular}
\end{center}
\end{table}

\section{Preliminary Result and Outlook}

The preliminary result presented in this paper 
is obtained from the full sample of
reconstructed $\kpppc$ decays collected by the experiment NA48/2 during 
 the data taking period of the year 2003. Three independent analyses,
 which agree within uncorrelated uncertainties,
 have been performed and averaged. The result is calculated separately
 for each of the four super-samples and then combined taking into
 account correlated systematic uncertainties
 (Table~\ref{tab:res}).

\begin{table}[t]
\caption{Measurement of $\Delta g$ in units of
  $10^{-4}$ in individual super-samples and the corresponding
  statistics. Only uncorrelated uncertainties,
  statistical uncertainty combined with the uncertainty from the 
  level 2 trigger
  efficiency, are quoted. \label{tab:res}}
\vspace{0.4cm}
\begin{center}
\begin{tabular}{|l|r|r||r|}
\hline
Super-sample & $\kpppp$ in $10^6$ & $\kpppm$ in $10^6$ & $\Delta g
\times 10^{4}$ \\ \hline
0     &  431 & 240 & $0.5 \pm 2.4$ \\
1     &  258 & 144 & $2.2 \pm 2.2$ \\
2     &  253 & 141 & $-3.0 \pm 2.5$ \\
3     &   95 &  53 & $-2.6 \pm 3.9$ \\ \hline
Total & 1036 & 577 & $-0.2 \pm 1.3$ \\
\hline
\end{tabular}
\end{center}
\end{table}

\begin{figure}
\begin{center}
\psfig{figure=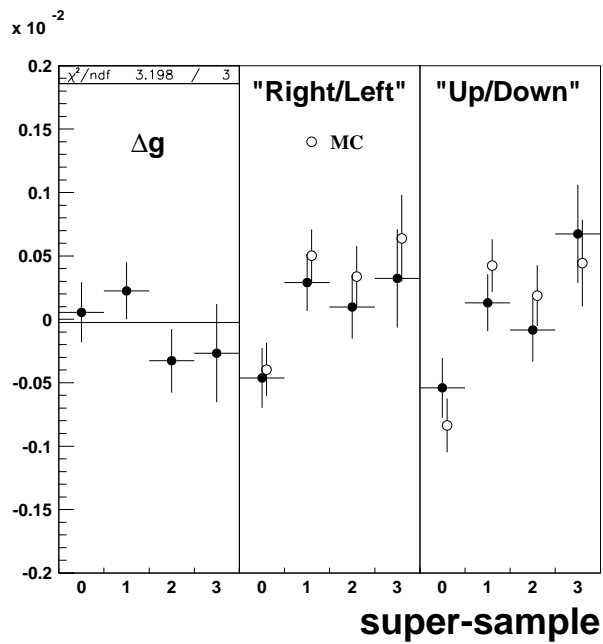,width=9cm}
\caption{Left: Measurement of $\Delta g$ in four super-samples. Middle,
  Right: Detector resp. beam-line asymmetries which cancel in the
  Eq.~\ref{eq:quadr} and their comparison to the Monte Carlo simulation.
\label{fig:result}}
\end{center}
\end{figure}

The measurement stability as a function of super-sample is shown in the
left part of the Fig.~\ref{fig:result}. All four measurements are
compatible with each other with $\chi^2/ndf = 3.2/3$. 
As a systematic check, in the middle part of this
figure, the quadruple ratio in Eq.~\ref{eq:quadr} is rearranged such
that instead of four $K^+/K^-$ ratios, four ratios of samples 
in which even pions are deflected to the right are divided by samples
with even pions deflected to the left in the spectrometer magnet.
In this case, the physical quantity $\Delta g$ cancels and the result is
expected to be equal to zero in the absence of any residual left-right
detector asymmetries. Similarly, the right part of the Fig.~\ref{fig:result}
reflects the asymmetry of the two beam paths. These
control asymmetries, which cancel at first order in Eq.~\ref{eq:quadr},
show that the cancellation of systematic
biases due to residual time variable imperfections in the apparatus is
at the level of few $10^{-4}$ and therefore second order effects are
negligible. Moreover, the comparison with Monte Carlo simulations
shows that all these apparatus asymmetries are understood in terms
of local inefficiencies and beam optics variations.

The combined preliminary result from all four super-samples is
\be
\Delta g = (-0.2 \pm 1.0_{stat.} \pm 0.9_{stat.(trig.)} \pm
0.9_{syst.})\times 10^{-4}
\ee
Converted to the asymmetry using the 
PDG value of the Dalitz slope $g$~\cite{pdg}:
\bea
\ag &=& (0.5 \pm 2.4_{stat.} \pm 2.1_{stat.(trig.)} \pm 2.1_{syst.})
\times 10^{-4} \\
    &=& (0.5 \pm 3.8)\times 10^{-4}
\eea

This result is compatible with no CP violation and with Standard
Model predictions but has more than an order of magnitude better 
precision than
similar previous measurements~\cite{exp}. Further improvements are expected in
future. The uncertainty due to the trigger can be
significantly improved by deeper study of 
inefficiencies as well as by emulating the trigger in 
Monte Carlo simulations. 
Also some of the systematic uncertainties can be improved by
more detailed studies of involved effects and by using data collected in the
period of 2004. These data are being analysed and will significantly
reduce the statistical uncertainty by more than doubling the total
sample.

NA48/2 has collected also about 200 millions of $\kpppn$ decays. This
decay mode is disfavored statistically due to its lower branching ratio
and lower acceptance. On the other hand, this disadvantage is
practically completely compensated by more favorable population of
the Dalitz plot leading to an expected statistical uncertainty
on $\ag$ comparable to that of the $\pppc$ decay mode. 

%\section*{Acknowledgments}

\section*{References}
\enlargethispage{28cm}

\end{document}